# OSC-MAC: DUTY CYCLE WITH MULTI HELPERS CT MODE WI-LEM TECHNOLOGY IN WIRELESS SENSOR NETWORKS

Mbida Mohamed and Ezzati Abdellah

Emerging Technologies Laboratory (VETE), Faculty of Sciences and Technology Hassan 1st University, Settat, MOROCCO

***ABSTRACT***

*Recently, Wireless Sensor Networks (WSNs) grow to be one of the dominant technology trends; new needs are continuously emerging and demanding more complex constraints in a duty cycle, such as extend the life time communication . The MAC layer plays a crucial role in these networks; it controls the communication module and manages the medium sharing. In this work we use OSC-MAC tackles combining with the performance of cooperative transmission (CT) in multi-hop WSN and the Wi-Lem technology*

## 1. INTRODUCTION

The Protocol OSC-MAC (On-demand Scheduling Cooperative Mac) is a protocol with a planning and request in order to solve the problems of energy of nodes transmitter, and gives as solution the integration of the technique of cooperation with the nearby nodes. In our work we integrate the Wi-Lem technology in order to delegate the nodes which have the higher energy and can be into the cooperation with the source, this solution brings an optimality of energy because the cooperative selection of nodes is applied in the Base station Wi-Lem, so we opted for 2 or more nodes which can be into the CT (cooperative transmission) according to the number of packets sending.
.
## 2. THE OSC-MAC PROTOCOL

As we know the life time of batteries is limited , what causes during the sending an imbalance energy in nearby nodes , and results an exhaustion power of the TRN (Transmitter root node) , to solve this problem we opt to initialize the technique by a phase of wakeup and decentralize the responsibility of sending's on the nearby nodes . Then according to the conception of nearby nodes which want to cooperate with these neighbour's and the others which refuse, during this work we are going to explain the state of art the request list of the cooperative nodes for a transmitter and the algorithm which engenders sending of packages by basing itself on the CT (cooperative transmission) RDV .  The cooperating nodes are neither on the same duty cycle nor are they in the same collision domain. We use orthogonal and pipelined duty-cycle scheduling, in part to reduce traffic contention, and devise a reservation-based wake-up scheme to bring cooperating nodes into temporary synchrony to support CT range extension.

The major sources of energy consumption inherent to MAC's include idle listening, overhearing and collision, besides data transmission and reception. Idle listening means that nodes keep listening to the channel while there are no incoming packets at all - a case that has not been taken care of in many MAC protocols such as IEEE 802.11 where in WIFI stations must listen for possible traffic. Notably, idle listening is disastrous in WSNs based on the fact that nodes in this mode consume the same magnitude of power as in receiving. Overhearing means that nodes decode packets that are destined to others. Collisions result in corrupted packets and the following MAC layer retransmissions consume extra energy. From the network perspective, albeit these factors taper off individual node's lifetime, the network lifetime is more critically limited by the energy holes formed around the sink leaving unused energy outside of the holes. Many authors have considered duty-cycle MAC protocols, which allow nodes to alternate between active and sleep modes. These protocols dramatically reduce the periods of idle listening and overhearing.

## 3. SELECTION OF NODES IN ENERGY FACTOR BY USING THE WI-LEM TECHNOLOGY

Instead of making a selection of nearby nodes in the transmitter which results an additional loss of energy, for that reason we migrate this operation in a Wi-Lem station with large domain of communication radio.

### 3.1. WI-Lem technology

LEM has just create the family of Wi-LEM (Wireless Local Energy Meter, wireless local meters of energy) to allow the measure and the remote surveillance of the energy consumption as the power of batteries nodes or the water as well as the temperature and the humidity. With these meters, the industrial and tertiary companies can reduce their consumption of energy and water as well as identify the points of improvement. On the whole range Wi-LEM, allowing a large domain of communication radio between the nodes of the network, compared to the previous generation.

### 3.2 The advantages of WI –Lem

Wi-LEM presents additional benefits compared to the traditional divisional meters in wsn:

- At the level of the electric cupboard

- At the level of the installation of the network and its exploitation.

The following figure (1) presents the operating Wi-Lem CT technology in wsn:

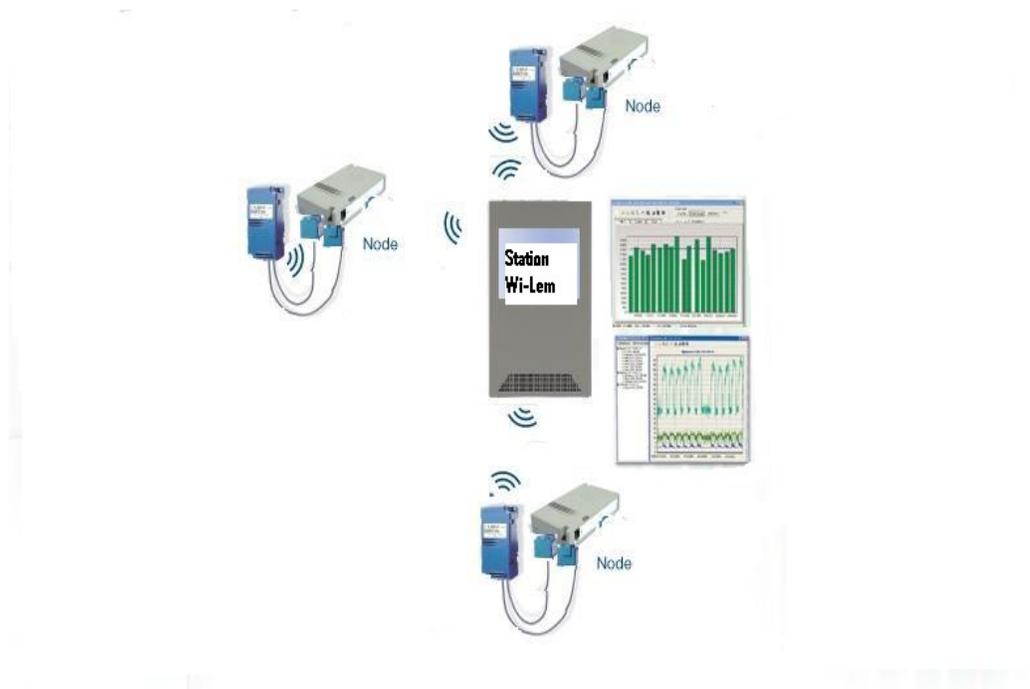

## 4. SELECTION OF THE NODES COOPERATION BY THE WI-LEM STATION

When a transmitter node wants to send its packets, first it transmit a packet request list of CTs nodes to the station Wi-Lem by using an algorithm of selection nearby nodes possible to operate in CTs periods, the algorithm (1) and the figure (2) illustrates this function :

**Algorithm 1 : allocation of the CTs in the nearby nodes**

**Input : List of the nearby knots organized has the order decreasing by
           percentage of energy : L
           The number of (octets) sent per packet : S
           Distance between transmitter and the next hop : D**
**Output : List of nearby nodes elected to transmit in CT phase    :
           LCT**

**For each li Є L do**
    **If Eli >= ( Elect * S) + ( Empl * S * D ²) ;**
    **LCT ←li ;**
    **Else Li ← li+1 ;**
    **End**
    **Send LCt () ;**
    **End**

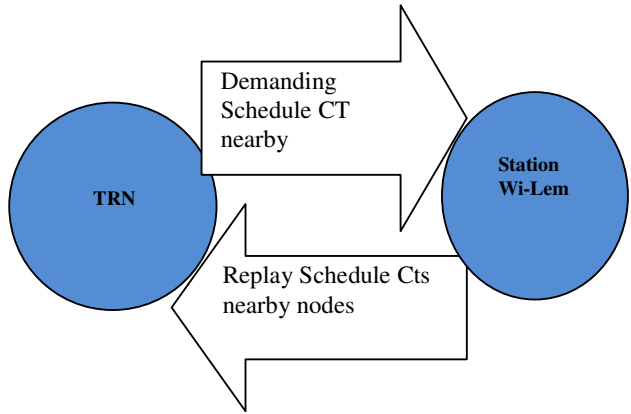

Figure 2: Diagram of the request /replay List CTs nearby

## 5. ELECTION OF THE NEARBY NODES WHICH WILL GO INTO THE CT MODE

Before the transmitter synchronizes its wakeup with the receiver as well as the nodes which want to run in CT Mode, the sending of packets is made according to report energy of nodes CT : { E ni ct (wi-lem) / i Є {1…n} , n : number of nearby node's in wsn } obtained by the measurement of energy nodes in station Wi-Lem, and also get the necessary energy to send N packets by node transmitter : { E niT / i Є { 1…t} , t : t ith node transmitter } , for this reason we develop The following algorithm (2) which explains the calculation of the necessary energy to transmit one packets of L bits :

| **Algorithm 2 : Calculation of energy transmission packets for L bits  E ni T** |
|---|
| **Input :  Emp-data : amplifying energy**<br>         **Efs         : free space energy**<br>**Output : Energy of sending one packet with L bits** |
| **if d is distance and do=sqrt (efs/emp);**<br>**then sending L bit data over d distance**<br>**if d>=do ;**<br>**Energy=energy-(l\*Etx+l\*emp\*(d^4));**<br>**else**<br>**Energy=energy-(l\*Etx+l\*efs\*(d^2));**<br>**End** |

As a reference of the ratio energy of nodes and the energy of transmission packet of L bits determine the number of helpers which are going to operate into the CT mode: the algorithm (3) describes the process of delegation nearby nodes CT.

International Journal of Wireless & Mobile Networks (IJWMN) Vol. 6, No. 2, April 2014

| Algorithm 3 :  Selection of Elected  nearby node CT |
|---|
| **Input :** Lists of the energy nodes which can enter in the CTs    RDV : E L (ct )<br>         Number of packets transmission : N<br>**Output :** List of elected nearby nodes : L ct-elu |
| **For each ni Є E L (ct ) do**<br>**If E ni Ct / (N\*E ni T)  >= 1**<br>**L Ct-elu ←ni ;**<br>**ni ←ni +1 ;**<br>**Else  ni ←ni +1 ;**<br>**Send L ct-elu() ;**<br>**End** |

## 6. ALLOCATION OF THE LISTS TRANSMISSIONS / RECEPTIONS PACKETS

In order to save the energy of every node sender we use a technique of Cooperative transmission which shows itself by the use of the nearby nodes according to its energy levels which run in routing mode of packets to the next hop, until the final Receiver. The  transmission of packets are initialized by a  phase of wake up  and allocation of the interval of sending's which divides in  no CT and CT intervals , in order to have   the transmitter node   and the cooperative nodes ( helpers ) are synchronized with the FR (final receiver) ,  the  figure 3 illustrate   a description  of this technique :

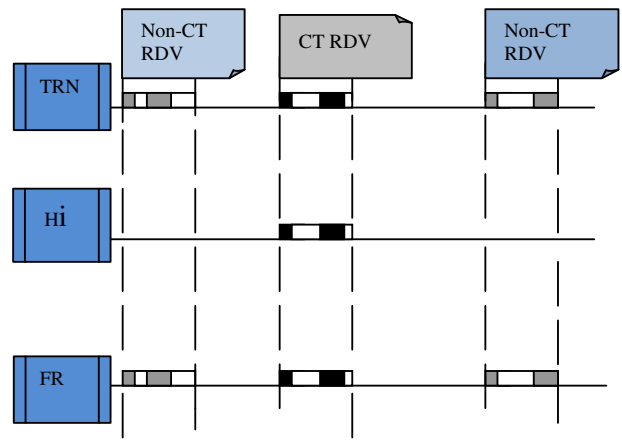

☐   Free period schedule
▨   CT RDV period schedule
■   No-CT RDV period schedule

**Hi (the helper's)** {the cooperative list nearby nodes / i Є {1 ... N}}
Figure 3: Diagram of allocation slot's time wake up for the CT and No-CT packets

The transmitter send a Super frame which defines the periods CT helpers and receiver in order to maintain the period of wake up, and to forward packets in the next hop, the leader Helper with the energy raised will send back a acknowledgement (ack) to inform the root transmitter node that the CT's meeting is reserved, the figure 4 describes this operative function. In no-Ct mode the sender node send directly a request in the next hop, up to the final receiver by indicating that it is going to transmit in a specific interval of time and the receiver will answer by a replay ack to inform it that the reservation of interval time is accepted or rejected.

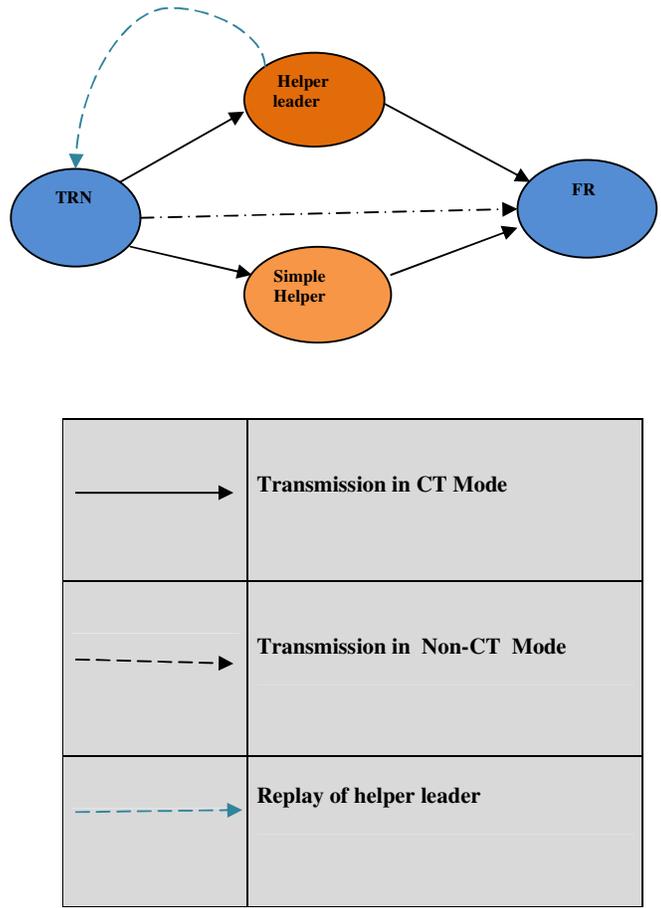

Figure 4: Diagram of transmission OSC-Mac by using CT Wi-Lem technology for 2 helpers

## 7. CONCLUSION

This paper propose the OSC-Mac Wi-Lem under exploitation of the cooperative transmission in goal as a technology to extend the life time of nodes by migrating the operations of classification energy's nearby nodes from the sender node to the base station Wi-Lem , and to list the nodes which are going probably to send into CTs periods .Indeed, it's used to choose the elected nearby nodes with higher power of the CTs period according to the number of packets sending, and simultaneously putting the other nearby nodes in a sleeping mode .

## Authors

Mbida Mohamed received the B.Sc. degree in networks and Information Systems, from Hassan 1 St University, Faculty of Sciences and Technology of Settat, Morocco, in2009 and M.Sc. degree in Network and Computer engineering from the Hassan 1 St University, Faculty of Sciences and Techniques (FSTS), Settat, Morocco, in 2012.Currently pursuing his PhD in Networks and Security Engineering at the Laboratory of Emerging Technologie (VETE) , from Hassan 1 st University, Faculty of Sciences and Technology of Settat, Morocco. His main research areas are how to use wireless sensor networks to secure and monitor mobile laboratories networks.

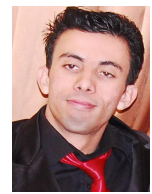

EZZATI Abdellah research Scientist in Faculty of Science and Technology in Morocco.He obtained his PHD in 1997 in Faculty of science in Rabat and member of the Computer commission in the same Faculty. Now is an associate professor in Hassan First University in Morocco and he is the Head of Bachelor of Computer Science .He participate to several project as the project Palmes which elaborate a Moroccan Education Certification.

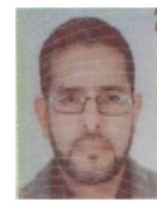